# Silicon-compatible ideal antiferroelectricity with large digital electromechanical responses enabled by thermal-strain domain engineering

Hao Xiong[1†], Huazhang Zhang[2,3†], Liang Shu[4†], Yangyang Si[1*], Jiaqi Liu[5], Chao Zhou[1], Rui Zhang[1], Jingxuan Li[1], Jinyang Li[1], Chhavi Rastogi[6], Hao Pan[7], Bin Xu[8], Er-Jia Guo[9], Yunlong Tang[5], Sujit Das[6], Philippe Ghosez[3], Qian Li[4], Jing-Feng Li[4], Zuhuang Chen[1*]

[1] State Key Laboratory of Precision Welding and Joining of Materials and Structures, School of Materials Science and Engineering, Harbin Institute of Technology, Shenzhen, 518055, China

[2] School of Physics and Mechanics, Wuhan University of Technology, Wuhan, 430070, China

[3] Theoretical Materials Physics, Q-MAT, University of Liège, B-4000 Sart-Tilman, Belgium

[4] State Key Laboratory of New Ceramic Materials, School of Materials Science and Engineering, Tsinghua University, Beijing, China.

[5] Shenyang National Laboratory for Materials Science, Institute of Metal Research, Chinese Academy of Sciences, Shenyang 110016, China

[6] Materials Research Centre, Indian Institute of Science, Bangalore, 560012, India

[7] State Key Laboratory of Advanced Waterproof Materials, Guangdong Provincial Key Laboratory of Nano-Micro Materials Research, School of Advanced Materials, Shenzhen Graduate School, Peking University, Shenzhen, 518055, China

[8] Jiangsu Key Laboratory of Frontier Material Physics and Devices, School of Physical Science and Technology, Soochow University, Suzhou 215006, China

[9] Beijing National Laboratory for Condensed Matter Physics and Institute of Physics, Chinese Academy of Sciences, Beijing 100190, China

[†] **These authors contributed equally to this work**

[*] **Corresponding authors:** zuhuang@hit.edu.cn; siyangyang@hit.edu.cn

**Antiferroelectrics exhibit reversible antipolar-polar transformations, offering a compelling platform for multiple functionalities in modern nanoelectronics, yet deterministic control of antiferroelectric domains and switching pathways remain elusive. Moreover, their integration with ubiquitous silicon-based electronic devices has been limited by the structural and chemical incompatibilities of conventional oxide platforms. Here, we convert the conventional drawback of thermal mismatch into a functional advantage and realize ideal antiferroelectricity in epitaxial $PbZrO_3$ thin films on silicon through thermal tensile-strain engineering, a strain regime unattainable on conventional perovskite substrates. Combined theoretical and experimental studies show that tensile strain stabilizes the $(004)_o$ domain, enabling a direct one-step switching, whereas compressive-strain-stabilized $(240)_o$ domains switch through intermediate ferrielectric states. The resulting films exhibit near-zero remanent polarization (~0.5μC/cm$^2$), square double hysteresis, nanosecond switching (~75ns), large reversible electrostrain (~0.6%) and robust operation windows. These findings provide key insights into domain-engineered ideal antiferroelectricity on silicon, opening a viable route toward high-performance antiferroelectric nano-electronic devices.**

## Introduction

Antiferroelectric (AFE) materials, which are characterized by reversible antipolar-to-polar transitions under the application or removal of an electric field, have attracted considerable attention for next-generation electronic devices [1], including high-density energy storage [2,3], electrocaloric cooling [4], thermal switching [5], negative capacitance transistors [6], and multi-state memories [7]. Beyond these functionalities, AFEs offer a unique electromechanical property: the field-induced phase transition is accompanied by a sharp volume expansion, producing a large "digital" electrostrain, a bistable on/off displacement with a constant step at a narrow threshold field. This contrasts fundamentally with the analog, continuous response of conventional piezoelectrics and is highly attractive for nanoelectromechanical systems (NEMS) that demand binary, digitally controlled actuation. However, translating this promise into a mainstream technology requires monolithic integration with silicon, which has been the cornerstone of modern nanoelectronics and a goal remained elusive due to two intertwined challenges.

First, at the device level, there is substantial chemical incompatibility and a structural mismatch between perovskite AFE oxides and silicon substrates, leading to a high density of defects [8]. More critically, the significantly lower thermal expansion coefficient of Si (~$2.6 \times 10^{-6}$ /K) compared to that of perovskite oxides (~8-9 $\times 10^{-6}$ /K) induces large strain during cooling after deposition, causing defects and even cracks that are usually viewed as a reliability nightmare [9]. Second, at the materials level, taking $PbZrO_3$ (PZO) as an example, it suffers from intrinsic structural complexity that is sensitive to external stimuli and boundary conditions [10–12]. With reduced dimensions, PZO thin films exhibit strong competition among multiple ferroic orders, including ferroelectric (FE), ferrielectric (FiE), and AFE phases [13–17]. These metastable polar structures in PZO usually result in the coexistence of FE, FiE, and AFE phases, which severely distort antiferroelectric behavior and degrade application performance [5,18].

On the other hand, although epitaxial strain has been widely used to stabilize specific domains and tune polar switching pathways in thin films with suitable lattice constants [19,20], this approach fails for materials with large lattice constants exceeding those of all commercially available perovskite substrates (≈3.70-4.02 Å), which remains a challenge that is particularly acute for AFEs. In the prototype AFE PZO, the bulk phase at room temperature is a distorted perovskite structure of *Pbam* orthorhombic symmetry with lattice constants $a_o$ = 5.888 Å, $b_o$ = 11.771 Å, and $c_o$ = 8.226 Å (subscript *o* denotes the orthorhombic index) [1]. The orthorhombic unit cell can alternatively be described as a simplified pseudocubic unit cell with $a_{pc} = b_{pc} \sim 4.161$ Å and $c_{pc} \sim 4.113$ Å (subscript *pc* denotes the pseudocubic index). The lattice parameters of PZO are much larger than the in-plane pseudocubic lattice parameters of commercially available perovskite oxide single crystal substrates, which range between ~3.70 and ~4.02 Å [1], rendering significant strain relaxation via misfit dislocations. As a result, the PZO films suffer from substantial interfacial defects as well as residual compressive strain from the perovskite oxide substrate, which is predicted to stabilize the FiE/FE phases and thus severely distort the AFE properties [21,22].

The limited choice of commercially available single-crystal substrates with sufficiently large lattice parameters has rendered tensile strain inaccessible in epitaxial $PbZrO_3$ thin films via conventional strain engineering. As a result, the ability to take full advantage of strain effects in AFE thin films can hardly be achieved through conventional epitaxy growth [18], hindering the deterministic control of AFE domain structure and correlated functionalities. Developing new strain-engineering approaches is therefore crucial not only for advancing the fundamental understanding of AFE switching mechanisms, but also for the functional design of AFE applications. In this context, the thermal tensile strain inherently generated during the cooling of Si-based heterostructures—typically viewed as a reliability hazard—presents a very challenging but compelling and complementary opportunity, as it simultaneously enables a large tensile strain regime and monolithic integration with CMOS-compatible platforms.

Here, we achieve ideal antiferroelectricity in epitaxial $PbZrO_3$ thin films integrated on silicon. By strategically leveraging thermally induced tensile strain, we realize deterministic control of the domain structure, stabilizing $(004)_o$-oriented (*i.e.*, *c*-domain) PZO films with superior AFE performance. Consequently, domain-mediated selective transition pathways are revealed by atomistic simulations and switching characterizations: the tensile strain-stabilized $(004)_o$-oriented PZO film exhibits fast, one-step polarization switching, whereas the compressive strain-stabilized $(240)_o$-oriented (*i.e.*, *a*-domain) PZO film undergoes a sluggish, multi-step transition. Specifically, the $(004)_o$-oriented film grown on Si enables ideal antiferroelectricity with near-zero remanent polarization, an ultrafast switching time of ~75 ns, a large digital electrostrain of ~ 0.6% that is highly applicable for nanoelectromechanical systems (NEMS), and excellent temperature and frequency stability. So, our work not only addresses the long-standing challenge of integrating high-performance AFE oxides thin film onto silicon, more broadly, it provides a scalable pathway toward AFE-based nanoelectronics and nanoelectromechanical systems that require fast, reproducible, and discrete displacement on a CMOS-compatible chip.

## Results

### Atomistic insight into the strain-stabilized antiferroelectricity

The energetics of PZO under biaxial strain and electric field were examined using a previously developed deep-learning interatomic potential [23]. This potential has been extensively validated (see Supplementary Table 1 for additional validation data relevant to the present study) and has been shown to retain near-first-principles accuracy while enabling efficient calculations for more complex structural configurations in larger supercells. Fig. 1a presents the energy comparison of several representative low-energy phases under different strain levels. The considered low-energy phases include (Fig. 1b): the *R*3*c* phase (a FE phase stabilized under electric field), the *Pbam* phase (the conventional AFE phase with a fourfold "↑↑↓↓" polarization pattern), the *Ima*2 phase (a FiE phase with a threefold "↑↑↓" polarization pattern, initially predicted by Aramberri et al. [24] and recently observed experimentally [13,25]), and the $Pmc2_1$ phase (a FiE phase with an eightfold "↑↑↑↑↓↑↑↓" polarization pattern identified by Burkovsky *et al.* in PZO films under electric field [17]). Except for the *R*3*c* phase, all other structures are orthorhombic and have two symmetry-inequivalent orientations under biaxial strain, *i.e.* the *ab*-oriented and *c*-oriented variants (corresponding to the $(240)_o$-oriented and $(004)_o$-oriented states and hereafter referred to as the *a*-domain and *c*-domain, respectively), which are schematically illustrated for the *Pbam* phase as an example in Fig. 1c and 1d, respectively.

The atomistic simulations show that the lowest-energy structure of PZO varies with strain. As shown in Fig. 1a, the *c*-oriented antipolar *Pbam* phase (*c*-*Pbam*) exhibits the lowest energy among all the examined structures within a specific range of tensile strain (from 0.5% to above 3.0%), whereas outside this range the lowest-energy structures are polar (*i.e.*, *R*3*c*, *ab*-*Ima*2, and *c*-*Ima*2). This result suggests that antiferroelectricity is favored at tensile strains, which is generally consistent with the first-principles calculations reported by Reyes-Lillo *et al.* [22], except that the present calculations consider additional FiE phases and predict a wider strain range over which *c*-*Pbam* is lowest in energy. It should be noted that the energy differences among all these phases are very small (on the order of a few meV per formula unit), so the strain range prediction should only be regarded as semiquantitative. For the orthorhombic structures, *i.e.*, *Pbam*, *Ima*2 and $Pmc2_1$, Fig. 1a also reveals a strain-controlled orientation preference. Specifically, the *ab*-oriented variants are energetically favored under compressive strain, while the *c*-oriented variants are favored under tensile strain. This behavior arises from the difference in lattice parameters along different crystallographic directions (Supplementary Table 1), where compressive strain favors the orientation with the shorter *c*-axis lying in-plane (*i.e.*, *a*-domain), while tensile strain favors the orientation with the shorter *c*-axis pointing out-of-plane (*i.e.*, *c*-domain).

A closer examination of the energies of the orthorhombic structures in the *ab*- and *c*-orientations reveals a strain-dependent competition between these phases. Under compressive strain, where the *ab*-orientation is preferred, the FiE *ab*-*Ima*2 and *ab*-$Pmc2_1$ phases progressively become more stable than the AFE *ab*-*Pbam* with increasing compression (Fig. 1e). Conversely, under tensile strain, where the *c*-orientation dominates, the FiE *c*-*Ima*2 and *c*-$Pmc2_1$ phases become less stable than *c*-*Pbam* as tensile strain increases (Fig. 1f). This competitive behavior also appears under an applied electric field, where the FiE *ab*-*Ima*2 and *ab*-$Pmc2_1$ exhibit a more rapid stabilization than *ab*-*Pbam* (Fig. 1g), due to their out-of-plane polarization component that couples directly to the applied electric field (Supplementary Fig. 1). In contrast, the *c*-oriented FiE phases, whose spontaneous polarization lies in-plane, do not show significant stabilization over *c*-*Pbam* as the electric field increases (Fig. 1h). Based on these energetic trends, it is plausible that under compressive strain the enhanced stability of the FiE phases may favor their involvement as intermediate states in the electric-field-induced *Pbam*-to-*R*3*c* transition, whereas under tensile strain, where such stabilization of the FiE phases is absent, a more direct, one step *Pbam*-to-*R*3*c* transition may be expected.

**Thermal strain induced *c*-domain selection in PZO thin films on Si**

Motivated by the above theoretical predictions, PZO films on both (001)-oriented STO and STO-buffered Si substrates, with $SrRuO_3$ as the bottom electrode, were synthesized by pulsed-laser deposition to compare the effect of strain on domain structure as well as on the AFE property. For the series of PZO films grown on STO and STO-buffered Si, the combination of atomic force microscopy (AFM) and X-ray diffraction revealed an atomically flat surface morphology with a root-mean-square roughness of ~650 pm (Fig. 2a) and high-quality epitaxial single-crystalline films without any secondary phase (Supplementary Fig. 2). Furthermore, it was found that the *a*-domain (with an out-of-plane lattice parameter of ~ 4.16 Å) is predominant when PZO is grown on STO. In sharp contrast, the *c*-domain (with an out-of-plane lattice parameter of ~ 4.11 Å) becomes predominant when PZO is grown on Si (Supplementary Fig. 2). Rocking curve measurements of a representative 230-nm-thick PZO film grown on Si yielded a small full width at half maximum (FWHM) value of 0.35° for the $(004)_o$ diffraction peak, validating its excellent crystallinity (Fig. 2a). These collective results demonstrate the successful growth of high-quality PZO films on Si. To confirm the epitaxial relationship of PZO on the Si substrate, φ-scans were performed on the $(103)_{pc}$ plane of PZO and the (115) plane of Si (Fig. 2b). The φ-scans show identical diffraction periodicity over a 0-360° rotation, confirming a 45° in-plane epitaxial rotation of PZO with respect to the Si substrate. This specific orientation is critical for mitigating the large lattice mismatch, thereby enabling a near-perfect registry between the perovskite lattice and half the diagonal length of the Si unit cell [26].

XRD peak fitting quantified the *a*/*c*-domain ratios in different samples by comparing the peak area and intensity (Supplementary Fig. 3) [27]. The result reveal that PZO films grown on Si substrates exhibit a higher proportion of *c*-domain compared to those grown on STO substrates (Supplementary Fig. 4). To further confirm the strain state and domain structure of films grown on different substrates, we conducted reciprocal space mapping (RSM) around the $(103)_{pc}$ reflection for PZO films grown on STO and Si (Fig. 2c, d and Supplementary Fig. 2). The RSM of PZO grown on STO shows typical out-of-plane superlattice reflections $(290)_o$ / $(450)_o$, originating from the antiparallel shift of $Pb^{2+}$ cations, indicating the presence of the *a*-domain with the AFE vector tilted by 45° from the out-of-plane direction. In comparison, no out-of-plane superlattice reflections are observed in the RSM of PZO grown on STO-buffered Si. Synchrotron-based in-plane RSM provides further insight. Two distinct in-plane $(372)_o$ superlattice reflections are observed around the main $(301)_{pc}$ reflection of PZO on Si (Fig. 2e). These peaks confirm the presence of the *c*-domain with in-plane antipolar ordering of $Pb^{2+}$ dipoles.

To confirm the strain state of PZO films on STO and STO-buffered Si, we analyzed the main $(103)_{pc}$ reflections of PZO films on both substrates, and found that the in-plane lattice parameter *b* of *a*-domain PZO on STO decreases from 4.161 Å (bulk value) to 4.154 Å. This small reduction indicates residual compressive strain imposed by the STO substrate. In contrast, *c*-domain PZO on Si exhibits increased in-plane lattice parameters (*b* of *c*-domain is as large as 4.209 Å) (Supplementary Table 2), demonstrating significant thermally induced tensile strain from the Si substrate. Such a large tensile strain is difficult to achieve using conventional lattice-matched perovskite oxide substrates. Importantly, this strain falls within the range predicted by our atomistic calculations to stabilize the AFE phase. These results directly highlight the critical role of thermal mismatch in generating large tensile strain and enabling domain control.

To characterize the domain structure at the atomic scale, cross-sectional STEM images of the PZO/SRO/STO/Si heterostructure were obtained, revealing clear interfaces in the stack (Fig. 2f). The atomic-resolution high-angle annular dark-field (HAADF)-STEM image of the PZO film clearly identifies well-ordered antiparallel dipoles with "↑↑↓↓" lattice of $Pb^{2+}$ displacement (Fig. 2g). Selected area electron diffraction (SAED) patterns show that the *a*-domain (Fig. 2h and Supplementary Fig 5) exhibits distinct 1/4-order superstructure spots along the (101) directions (as indicated by the arrows in Fig. 2h), confirming a quadrupled periodicity consistent with "↑↑↓↓" antiferroelectric ordering. In contrast, the SAED of the *c*-domain (Fig. 2i) shows no such superstructure reflections mainly due to the antipolar dipole order lying in-plane [28], which is fully consistent with the RSM results.

**Ideal antiferroelectricity in *c*-domain PZO films on Si**

To evaluate the AFE property of films on different substrates, the polarization switching characteristics were evaluated (Fig. 3) [14,29]. For PZO films deposited on the STO substrate, which are dominated by *a*-domain (Fig. 3a), clear multi-step switching was observed. This multi-step switching proceeds through intermediate FiE states, yielding up to ten switching current peaks and high remanent polarization ($2P_r$ as large as 7.02 μC/cm$^2$), consistent with previous work [10,11,18]. In contrast, *c*-domain PZO thin films deposited on Si substrates exhibit ideal antiferroelectricity with square double hysteresis loops (Fig. 3b). The double hysteresis loops show nearly zero remanent polarization ($2P_r$ as low as 1.06 μC/cm²), a steep slope at the critical electric field and no observable transient state during switching, as indicated by the four sharp switching current peaks. The C-V characteristics further corroborate the distinct switching behaviors. While the *a*-domain-dominant film exhibits multiple dielectric peaks, reflecting a multi-step phase transition pathway, the *c*-domain-dominant film shows a typical "double butterfly" shape with four sharp dielectric peaks consistent with abrupt switching (Fig. 3c). This contrast highlights the strong coupling between domain symmetry and the transition mechanism. The dielectric loss mirrors the behavior of the dielectric constant, further supporting this interpretation. In addition, the low dielectric loss (~0.03) and minimal leakage current (~$10^{-9}$ A cm$^{-2}$) confirm the excellent insulating characteristics and high crystalline quality of the films (Supplementary Fig. 7) [10,30]. To quantify the improvement with increasing *c*-domain fraction, the AFE superiority factor $\eta = (P_{max} - P_r)/P_r$ was evaluated [18]. As summarized in Supplementary Fig. 8, the *c*-domain PZO film on Si substrates achieves an AFE superior factor of ~79.4—competitive among both lead-based [2,11,31–35] and lead-free films [36–40]—demonstrating nearly ideal AFE characteristics in a Si-integrated platform (Fig. 3d).

While such Si-integrated AFE films exhibit excellent ambient performance, for practical applications, they must also demonstrate robustness under various operating conditions and modes. Previous studies have confirmed that, as temperature decreases, emergent ferroelectricity develops, accompanied by increased remanent polarization [14,41]. This evolution distorts the intrinsic AFE behavior and compromises its operational stability for device applications. To assess these critical attributes in the Si-integrated PZO films, the temperature- and frequency-dependent stability was systematically evaluated. Notably, *a*-domain-dominant PZO films on STO substrates present distinct FiE characteristics and a significant increase in remanent polarization with decreasing temperature, which severely undermines device stability (Fig. 4a). In sharp contrast, for *c*-domain-dominant PZO on Si, the antiferroelectric double loops remained well-defined throughout a wide temperature range from 77 K to 300 K, with the AFE–FE transition field decreasing as temperature increased. Notably, the remanent polarization remains nearly negligible from 77 K to 300 K (Fig. 4e), and the AFE phase persists up to 473 K (Supplementary Fig. 10), demonstrating the superior thermal stability of the *c*-domain–dominant PZO films compared with *a*-domain-dominant PZO films. Frequency-dependent dynamic hysteresis measurements (50 Hz to 100 kHz; Fig. 4d) show nearly ideal antiferroelectric loops, with stable saturation polarization and negligible remanence across the entire frequency range. These results confirm the robust and stable antiferroelectric behavior of the films on Si substrates under varying operating conditions.

**Enhanced switching kinetics and large digital electrostrain**

Apart from the dynamic hysteresis loops and C-V measurements, pulse measurements were further performed under identical electric fields to evaluate the phase transition kinetics of PZO films with different domain structures. Evidently, the *a*-domain-dominant PZO film on the STO substrate undergoes a multi-step transition from AFE to FE state, which is due to the FiE intermediate phases induced by the electric field [18]. In contrast, the *c*-domain-dominant PZO film on the Si substrate undergoes single-step switching without any intermediate phases (Fig. 4e and Supplementary Fig. 11). The complete switching time of the *a*-domain-dominant PZO film was further extracted from the time-resolved polarization switching process and measured to be ~ 490 ns. By contrast, *c*-domain-dominant PZO exhibits single-step switching with faster kinetics, achieving full polarization reversal within just ~75 ns, more than six times faster than the *a*-domain-dominant film. The results reveal a clear trend: a higher *c*-domain content leads to a faster AFE-FE phase transition (Supplementary Fig. 12). This more-than-sixfold acceleration in switching speed, coupled with the potential to reach the sub-nanosecond regime through electrode geometric scaling [42], underscores the transformative potential of domain engineering. This ultrafast switching behavior on a CMOS-compatible platform could be critical for high-speed memory and logic devices.

To model the switching characteristics of differently oriented AFE domains for practical applications, two established models: the Kolmogorov-Avrami-Ishibashi (KAI) [43] and the Nucleation-Limited Switching (NLS) [44] models, were employed. Both models were employed to simulate the switching kinetics of PZO films with varying *c*-domain fractions (Supplementary Fig. 13). For samples with a high portion of *a*-domain or *c*-domain, both models described the experimental data well, while for the sample with 57.1% *c*-domain fraction, the NLS model provided a better fit (Supplementary Fig. 13). This aligns well with the fact that the NLS model is more suitable for polycrystalline or ceramic films with structural disorder, while the KAI model describes homogeneous switching in single-crystal-like films [45]. The intermediate *a/c*-domain ratio likely introduces structural heterogeneity resembling a polycrystalline state, making the NLS more appropriate.

The ultrafast and sharp field-driven structural phase transition leads to an abrupt physical change that differs from that in conventional FE/piezoelectric materials, where the physical properties (e.g., piezoelectric responses) usually change continuously with applied electric field (referred to here as an analog-type response). For PZO with ideal antiferroelectricity, a digital-type electromechanical response, where a large volume change during transition could be obtained [46]. Specifically, the AFE phase ideally generates negligible displacement before the critical field. Upon crossing the critical field, the phase transition leads to an abrupt increase in displacement. This results in a strongly nonlinear, field-dependent electromechanical response—characteristic of digital-type electrostrain. While such digital electrostrain has been previously achieved on (111)-oriented epitaxial films grown on single-crystal perovskite substrates [18], the integration with silicon-based platforms is essential for compatibility with mainstream nanoelectronics. Based on the domain-engineered ultrafast one-step polarization switching, we investigated the electric-field-induced electromechanical response of the PZO films grown on Si. Using a laser Doppler vibrometer (LDV) [47], we find that the electromechanical response remains extremely low below 100 pm at low electric fields (i.e., below the AFE-FE transition field) (Fig. 5a). When the field exceeds the critical value, a sharp jump in displacement to over 1400 pm occurs, rendering a giant digital-type electrostrain of ~0.6% (Fig 5a, b). Unlike *a*-domain-dominant PZO films,  which exhibit an analog-type near-linear electrostrain with increasing electric field, resembling that of conventional ferroelectrics/piezoelectrics [18,40], this large digital electrostrain occurs within a narrow voltage window (Fig. 5c), directly mirroring the AFE-FE phase transition and the ultrafast switching kinetics of the *c*-domain-dominant PZO films, thereby enabling well-defined OFF/ON displacement states. Furthermore, the favorable frequency stability inherited from the nanosecond-scale dynamics (Fig. 5d), combined with integration on Si substrates, unlocks practical NEMS applications that demand fast, reproducible, and digitally controlled displacement.

**Conclusion**

In summary, we demonstrate the successful integration of ideal antiferroelectricity on silicon in antiferroelectric $PbZrO_3$ (PZO) thin films. By strategically harnessing thermally induced strain, we achieved deterministic control of domain structures in epitaxial PZO thin films, with the $(004)_o$-oriented domain exhibiting direct one-step switching, in contrast to the multistep switching of the $(240)_o$-oriented domain. This domain engineering resulted in near-zero remanent polarization, an ultrafast switching time of ~75 ns, and a large digital electrostrain of ~ 0.6%. Combined atomistic calculations and experimental characterizations confirmed that $(004)_o$-oriented domain-dominant films undergo a direct single-step switching pathway, in stark contrast to the multi-step process observed in their $(240)_o$-oriented domain-dominant counterparts. Moreover, the $(004)_o$-oriented films exhibit exceptional stability across a broad temperature range (77-473 K) and frequency range (50 Hz-100 kHz), underscoring their robustness for practical applications. This work not only provides profound insights into domain-mediated antiferroelectric switching but also establishes a robust and scalable pathway for the monolithic integration of high-performance functional oxides into mainstream Si-based nanoelectronics.

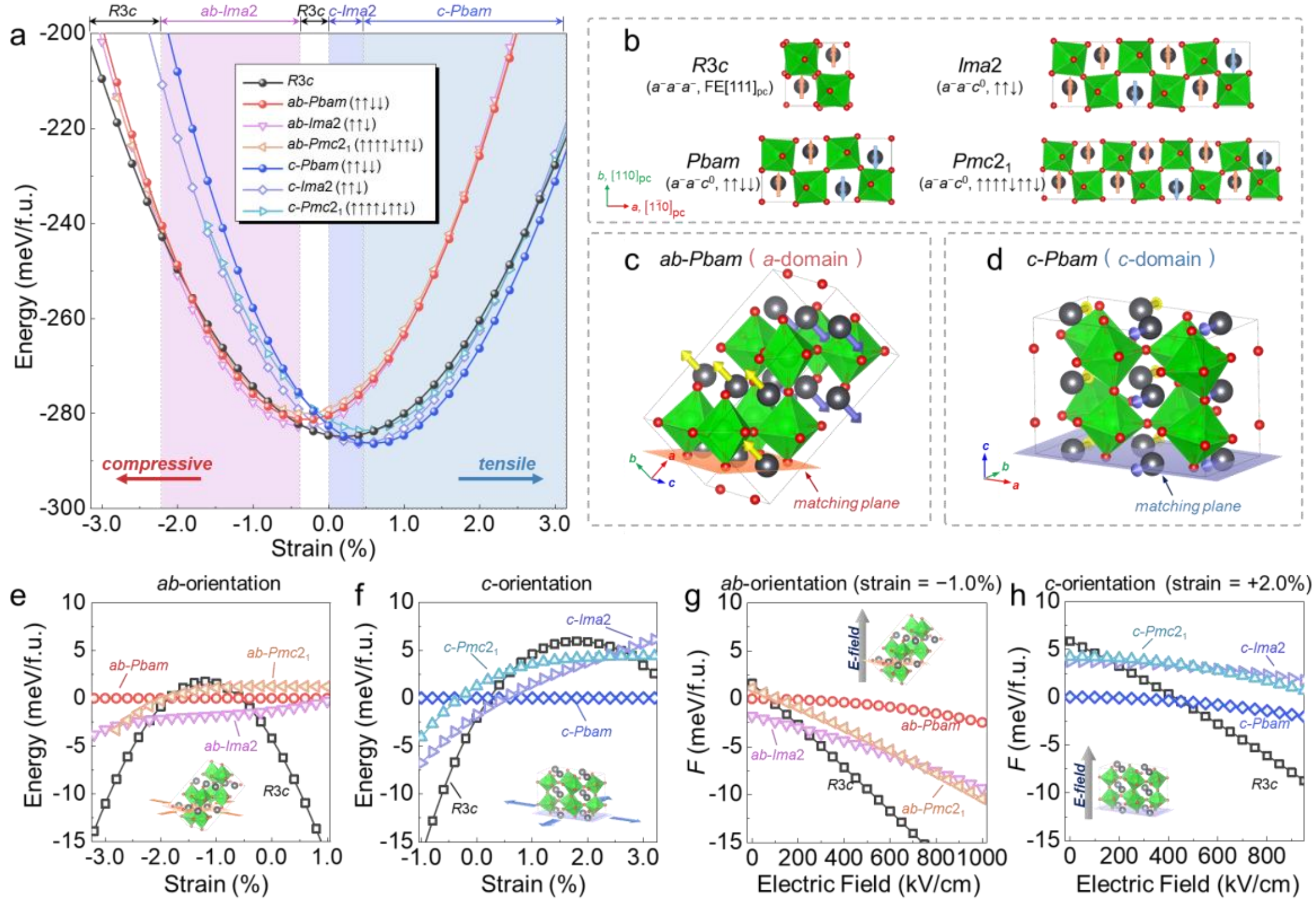

**Fig. 1 Energetics of PZO under biaxial strain and electric field. (a)** Energy as a function of biaxial strain for several representative low-energy phases, calculated using the deep-learning interatomic potential. Strain is defined relative to the reference lattice parameter $a_0$ = 4.133 Å (the cube root of the volume per formula unit of the *Pbam* phase), and energies are reported relative to the cubic phase. The energy of the *R*3*c* phase has been shifted upward by 5.0 meV/f.u. to correct its underestimation by the deep-learning interatomic potential (see Supplementary Table 1). The lowest energy phases at different strain ranges are indicated at the top of panel (a). **(b)** Schematics of the polarization patterns of the four low-energy phases considered: ferroelectric *R*3*c*, antiferroelectric *Pbam*, and ferrielectric *Ima*2 and $Pmc2_1$. **(c, d)** The two symmetry-inequivalent orientations for orthorhombic structures under biaxial strain, illustrated using the *Pbam* phase: (c) *ab-Pbam* and (d) *c-Pbam*, where the prefixes "*ab*-" and "*c*-" denote the orientation. Note that the biaxial strain lowers the space group symmetry of the *R*3*c* phase (to *Cc*) and the *ab*-oriented orthorhombic phases (*ab-Pbam* is lowered to *P*2/*m*, *ab-Ima*2 is lowered to *Cm*, and *ab*-$Pmc2_1$ is lowered to *Pm*), whereas the symmetry of the *c*-oriented orthorhombic phases are preserved. **(e, f)** Energy comparisons among different phases for (e) the *ab*-orientation and

(f) the *c*-orientation, with the *Pbam* phase as the energy reference in the respective orientations. **(g, h)** Stability of different phases under an electric field for (g) the *ab*-orientation under compressive strain of −1.0%, and (h) the *c*-orientation under tensile strain of +2.0%, evaluated using the energy functional $F$, defined by $F = U - \Omega(\mathbf{E} \cdot \mathbf{P})$, where $U$ is the total energy, $\Omega$ is the unit cell volume, $\mathbf{E}$ is the electric field, and $\mathbf{P}$ is the polarization.

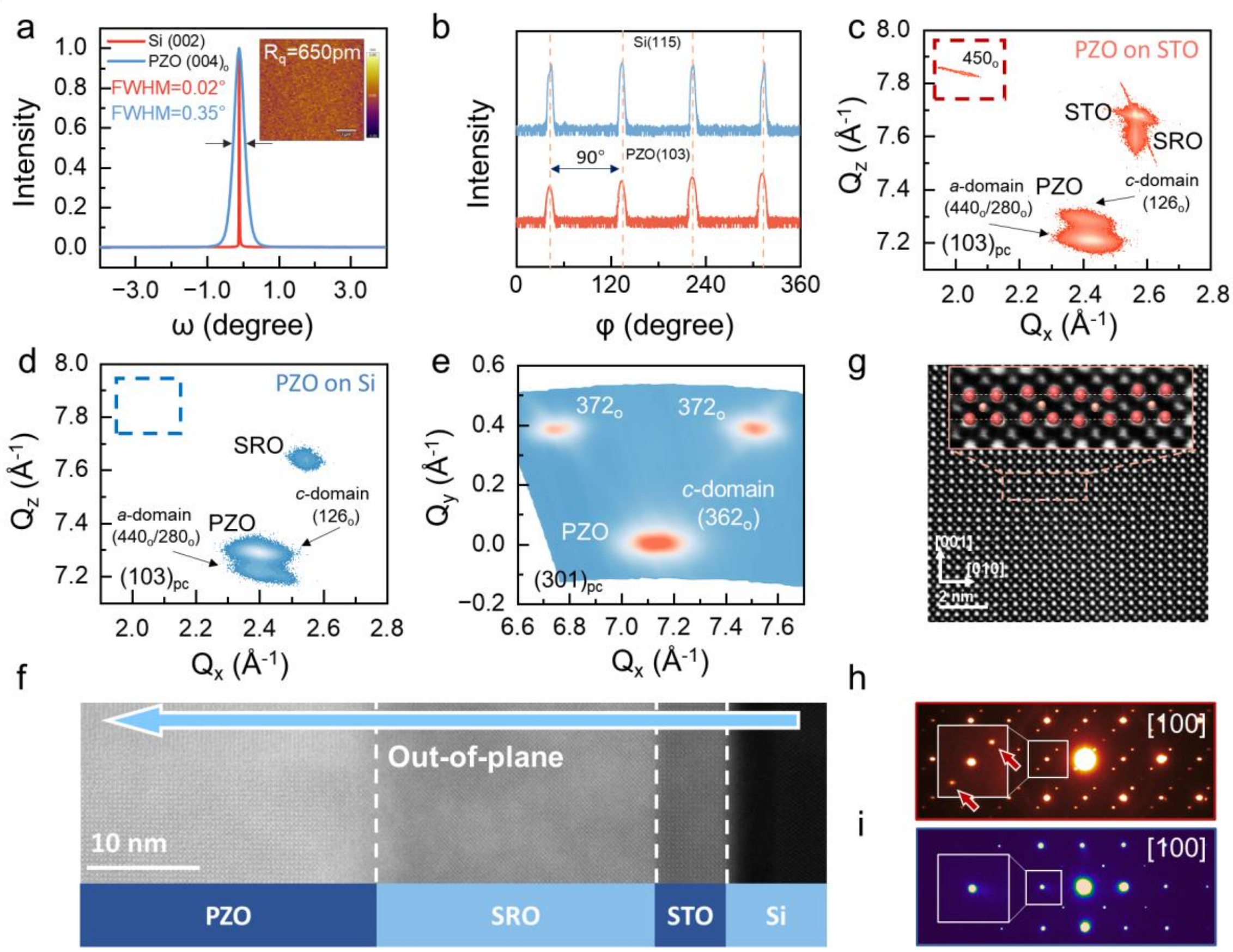

**Fig. 2** **Structural characteristics of PZO thin films grown on Si and STO substrates.** **(a)** Rocking curve around the $(004)_o$ reflection of PZO (230 nm)/SRO/STO-Si heterostructure (Inset: atomic force microscopy topographic image). **(b)** φ-scans of PZO $(103)_{pc}$ and Si (115) for the PZO/SRO/STO/Si heterostructure. **(c)** Out-of-plane RSM around the $(440)_o$ / $(280)_o$ reflection of a PZO film grown on SRO-buffered STO substrate, showing the presence of $(450)_o$ superlattice reflection spot. **(d)** Out-of-plane RSM around the $(126)_o$ reflection of the PZO/SRO/STO/Si heterostructure, showing the absence of a superlattice reflection spot. **(e)** Synchrotron-based in-plane RSM around the $(362)_o$ reflection of the PZO/SRO/STO/Si heterostructure, showing the characteristic $(372)_o$ superlattice reflection spots, which correspond to in-plane antiparallel cation displacements. **(f)** Low-magnified cross-sectional image of the PZO/SRO/STO/Si heterostructure. **(g)** The high-resolution HAADF-STEM cross-sectional image of the *a*-domain. **(h, i)** SAED patterns of the PZO on STO (*a*-domain) and Si (*c*-domain) substrates respectively; the arrows indicate the 1/4[110] superstructure reflections from antiparallel cation displacement (along the [100] zone axis).

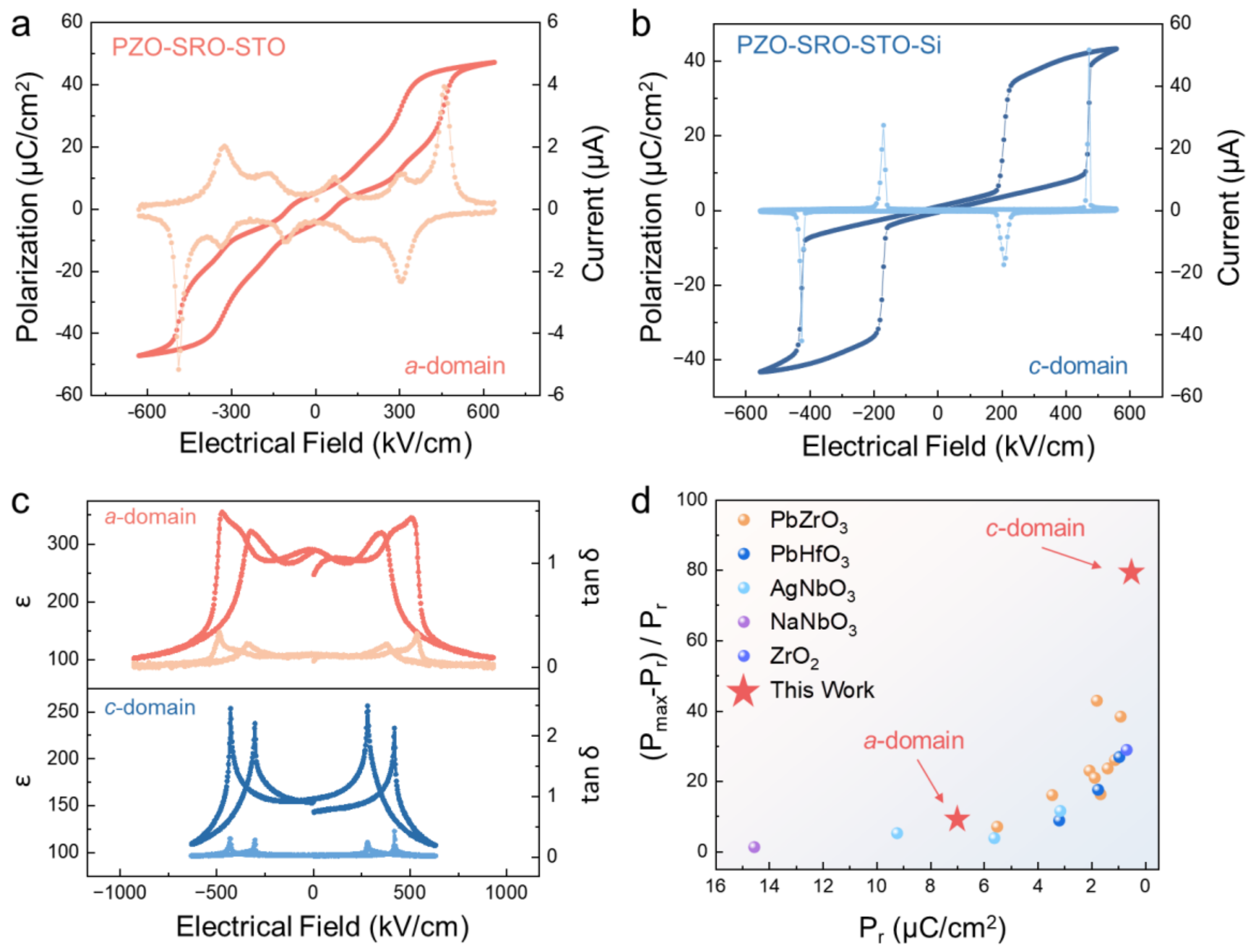

**Fig. 3 Electrical characteristics of antiferroelectricity in *a*-domain and *c*-domain PZO.** **(a, b)** P-E loops with switching current peaks for *a*-domain PZO/SRO/STO and *c*-domain PZO/SRO/STO/Si heterostructures. **(c)** Capacitance-voltage (C-V) characteristics, showing the relative permittivity ($\varepsilon_r$) and dielectric loss (tan δ), for *a/c*-domain-dominant PZO films. **(d)** Comparison of the antiferroelectric superior factors achieved in this work for the high *c*-domain-dominant PZO film against those of other representative antiferroelectric material systems previously reported (all films were deposited on (001)-oriented substrates).

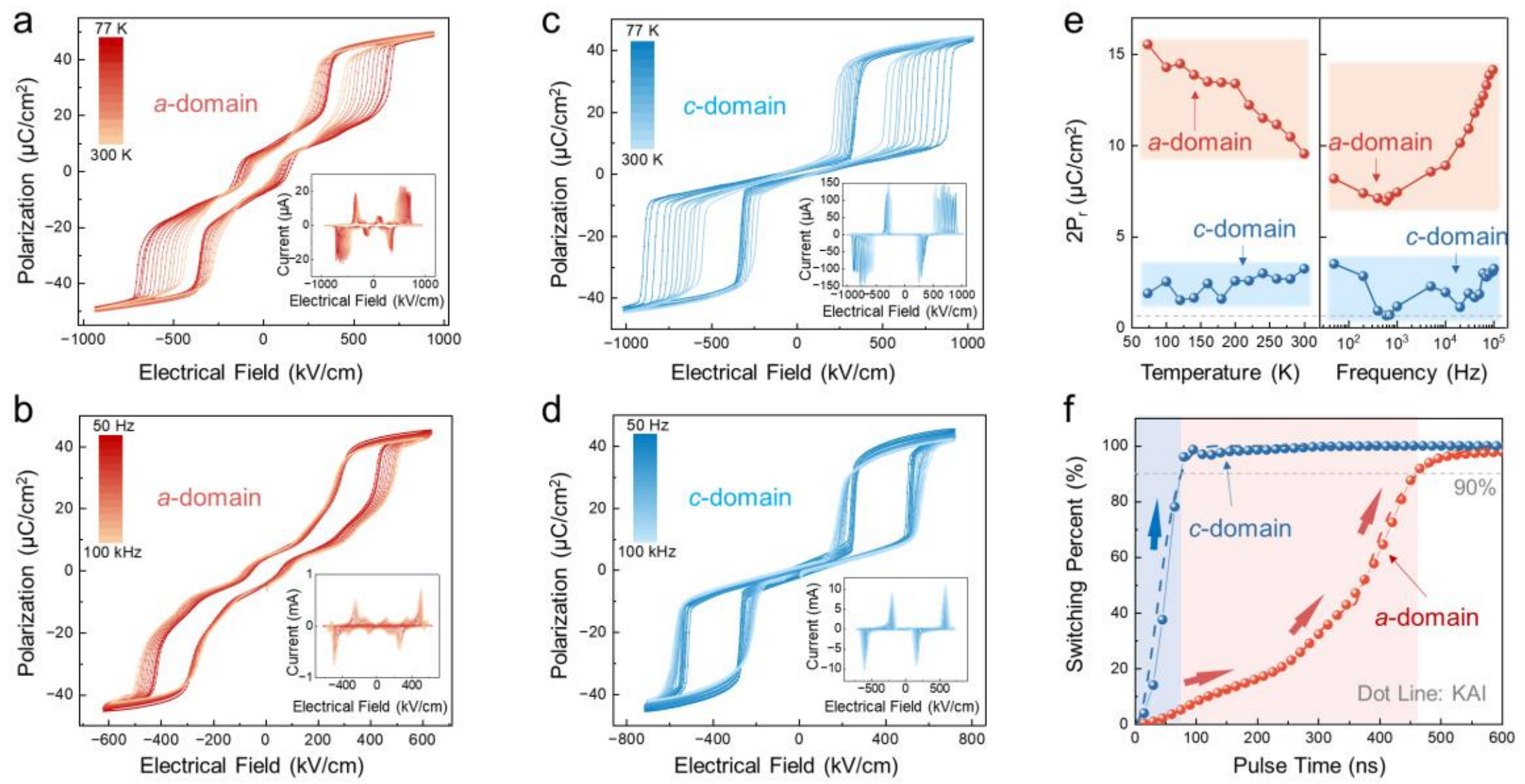

**Fig. 4 Stability of antiferroelectric property for different conditions and switching characteristics of antiferroelectricity in *a*-domain and *c*-domain PZO films. (a, b)** Evolution of the double hysteresis loops of *a*-domain-dominant PZO measured over a temperature range of 77-300 K and a frequency range of 50 Hz-100 kHz, respectively. **(c, d)** Evolution of the double hysteresis loops of *c*-domain-dominant PZO measured over a temperature range of 77-300 K and a frequency range of 50 Hz-100 kHz, respectively. **(e)** Statistical summary of the remanent polarization ($2\mathbf{P_r}$) from the hysteresis loops of (a, b) and (c, d) as functions of temperature and measurement frequency, respectively. **(f)** Switching kinetics (dotted lines) and fits using the Kolmogorov-Avrami-Ishibashi (KAI) model (dashed lines) of *a/c*-domain-dominant PZO. The switching time is defined as the time at which the polarization reaches 90% of its saturation value [42].

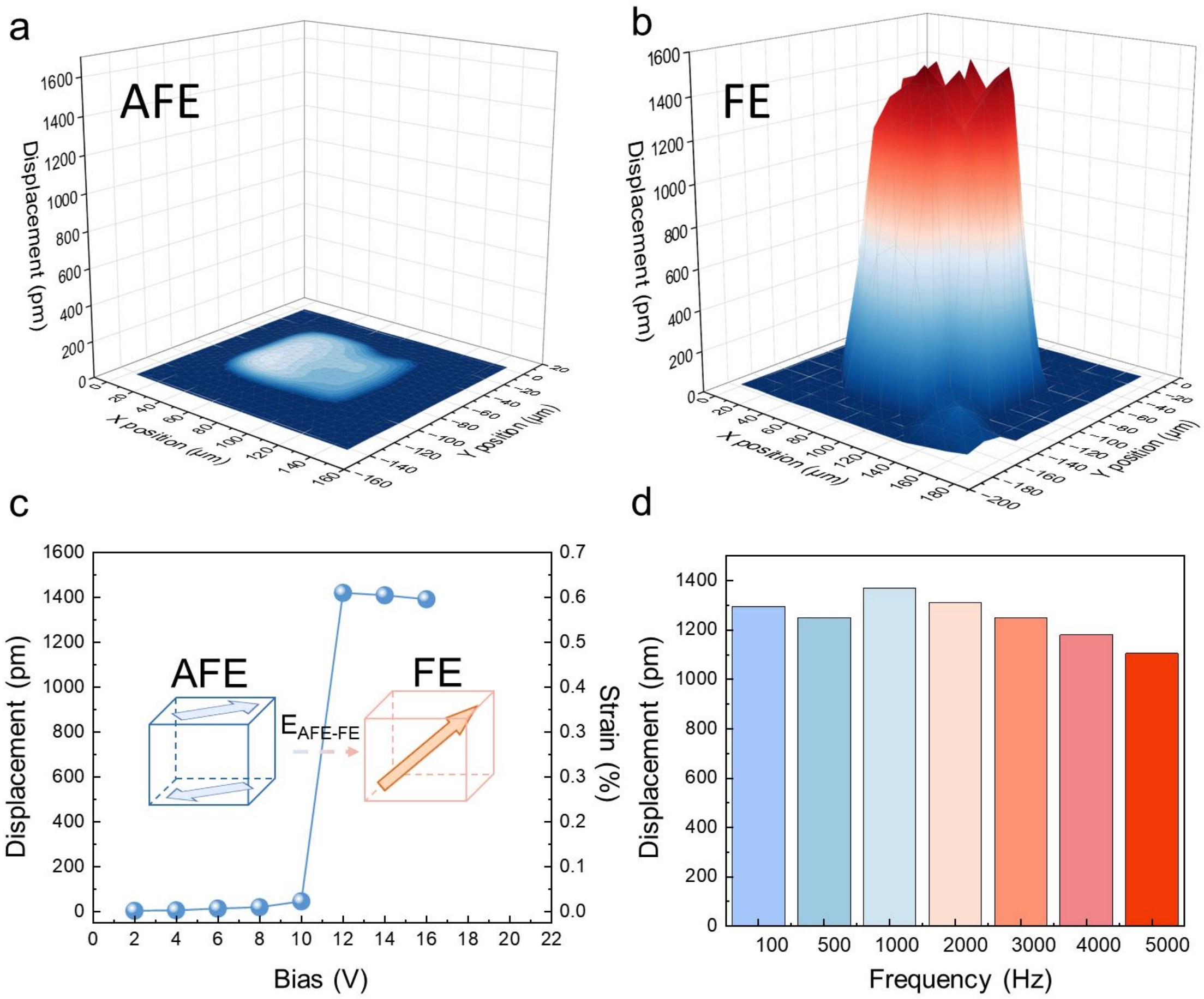

**Fig. 5 Electromechanical responses during AFE-to-FE phase transition of *c*-domain PZO films grown on silicon.** (**a, b**) Three-dimensional maps of surface displacements under applied electric field before and after AFE-to-FE transition, respectively. (**c**) Digital displacements response under different electric fields. (**d**) Frequency dependence of the displacements during the AFE-FE phase transition.

**ACKNOWLEDGMENTS**

The authors acknowledge the support from the BL02U2 of Shanghai Synchrotron Radiation Facility (SSRF).

**FUNDING**

This work was supported by National Natural Science Foundation of China (Grant Nos. 52525209, 52372105 and 92477129) and the Guangdong Basic and Applied Basic Research Foundation (Grant No. 2024B1515120010). Z.H.C. acknowledges the financial support for Outstanding scientific and technological innovation Talents Training Fund in Shenzhen and "the Fundamental Research Funds for the Central Universities" (Grant No. 2024FRFK03012). Y.L.T. has been supported by National Natural Science Foundation of China (Grant No. U24A2013). B.X. acknowledges financial support from National Natural Science Foundation of China (Grant No. 12574101).

**AUTHOR CONTRIBUTIONS**

Z.H.C. conceived and supervised this study; H.X. fabricated the films, performed the XRD, AFM and electrical measurements; H.Z.Z. and B.X. performed the atomistic calculations; L.S performed electromechanical response measurement; J.Q.L and Y.L.T. performed the STEM characterizations; H.X., H.Z.Z, Y.Y.S and Z.H.C. wrote the original draft; Z.H.C., Y.Y.S, H.X., H.Z.Z., L.S., J.Q.L., C.Z., R.Z., J.X.L, J.Y.L., B.X., Y.L.T., S.D., P.G., Q.L. and J.F.L. discussed the data and edited the manuscript. All authors commented on the manuscript.

**Competing interests**

The authors declare that they have no competing interests.

**Additional information**

Correspondence and requests for material should be addressed to Zuhuang Chen (zuhuang@hit.edu.cn).